**A new spin on spinning your samples: balancing rotors in a non-trivial manner**

Oleg E. Peil[1] and Vasili Hauryliuk[2]

[1]Department of Physics, Uppsala University, Box 530, S–751 21, Uppsala, Sweden

[2]University of Tartu, Institute of Technology, Nooruse St. 1, room 425, 50411 Tartu, Estonia

Corresponding author: Vasili Hauryliuk, vasili.hauryliuk@ut.ee

Due to a lack of coherent analysis, many common practices of humankind preserve low-efficient procedures. Balancing tubes during centrifugation exemplifies such a problem in laboratory practice. Using combination of symmetry group theory and genetic algorithm methodology we demonstrate that there is an array of surprisingly non-trivial algorithms for going about this procedure.

The common approach is to counter-balance tubes in such a way that the tubes are put pair-wise opposite each other. Evidently, this algorithm necessitates than an even amount of tubes are to be subjected to centrifugation. This is usually achieved by addition of one more "blank" tube with water in case the initial amount is odd. Apart from this technique, if the total amount of tubes is divisible by three, a tube constellation with $C_3$ symmetry rotational axis in the center of rotor is widely practiced. No other methods of rotor balance have gained popularity to date, thus limiting researchers' flexibility. Apparently, this situation can be improved by the development of proper auxiliary tools. To simplify the presentation, we consider a commonly used 30-slot rotor. If the notion of symmetry of a configuration is restricted to invariance under rotation by angle n·2π/30, where n = 2, 3, or 5 is a prime divisor of the total number of slots in the rotor, a class of solutions can be obtained by superposing configurations representing point groups $C_n$ [1]. All possible solutions representing a superposition of these symmetric configurations (further referred to as "symmetric solutions") are, therefore, generated by decomposing a desired amount of tubes into a sum of numbers 2, 3 and 5 with some restrictions due to the "exclusion principle": two tubes can not occupy the same hole. The simplest way to balance an odd amount of tubes this way is to place according to the configuration M·$C_2$+$C_3$ (Fig. 1A). If the number of tubes exceeds half of rotor slots, it is advisable to treat the problem as balancing *holes* instead of balancing *tubes*, simplifying the problem (Fig. 1B).

Unfortunately, this analysis is limited only to a certain sub-class of solutions that does not embrace the whole variety of balanced configurations. To solve a problem of finding all possible balanced configurations, we developed a genetic algorithm (GA) [2, 3] method designed to deal with problems of optimization over a complicated phase space with many local minima. Numerical simulations revealed a large number of non-symmetric, i.e. lacking *rotational* symmetry, configurations (exemplified on Fig. 1C) which can find their application in demonstration of superiority of senior researchers' experience, as compared with students', in handling lab equipment.

After developing the methodology, we turned to analysis of the readiness of the scientific community for the radical changes in the rotor balancing habits. For that matter, we conducted a social survey with 27 participants in Sweden, Estonia, US and Russia. Three arrangements of eppendorf tubes in $C_{30}$ rotor (Fig. 1A-C) were demonstrated to participants, and participants where asked whether arrangements are balanced. Combination of $C_2$ and $C_3$ symmetries composed of tubes was successfully recognized as balanced by roughly half of the participants (52 %) (Fig. 1D). The same combination but composed of "holes" was recognized by a smaller portion (32 %), apparently manifesting the fact that people concentrate attention on objects rather than on their absence. Finally, non-symmetric arrangement (Fig. 1C) was recognized as balanced by 17% of researchers. Some of these were actually calculating moment of inertia, i.e. were coming to solution knowingly, the rest where basically guessing. The latter should be banished from laboratory practice, since these people are ready to make dangerous decisions without actual understanding of the case, which renders them extremely dangerous in the laboratory settings.

**Figure legend:**

A. Balanced arrangement with 3 tubes forming $C_3$ symmetric ensemble and 4 pair-wise balanced tubes
B. The same as A, but holes instead of tubes are balanced
C. Non-symmetric balanced arrangement of 7 tubes in $C_{30}$ rotor
D. Results of survey on balance recognition. Denotation below bars corresponds to arrangements depicted on panels A-C
E. Distribution of symmetric and non-symmetric balanced arrangements for different numbers of tubes placed in the 30-slot rotor.

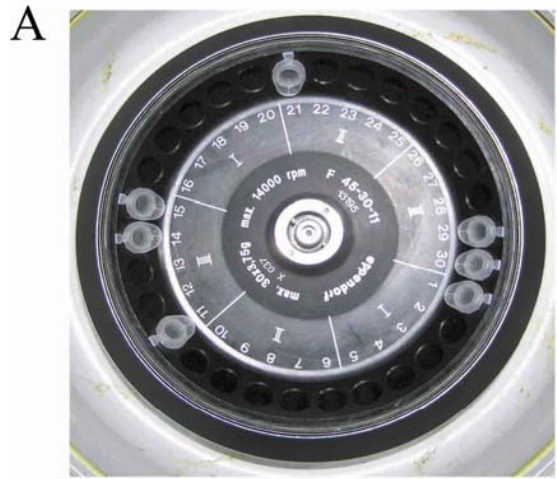
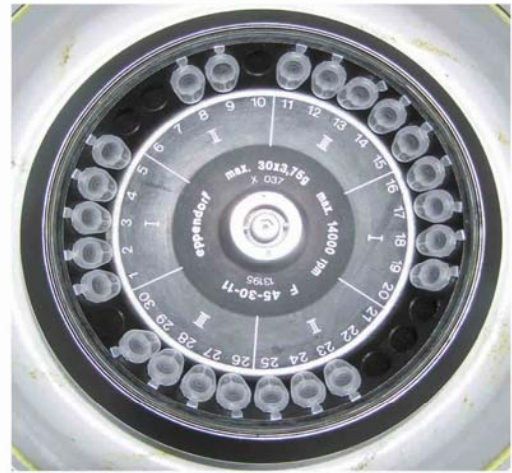
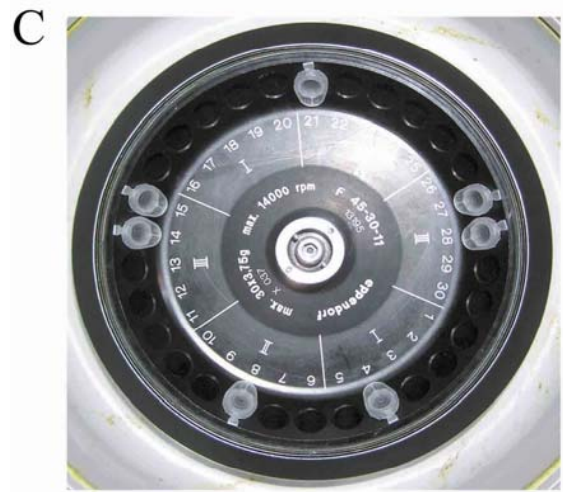
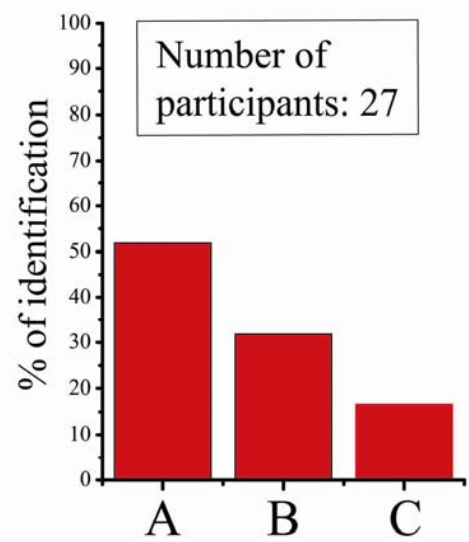
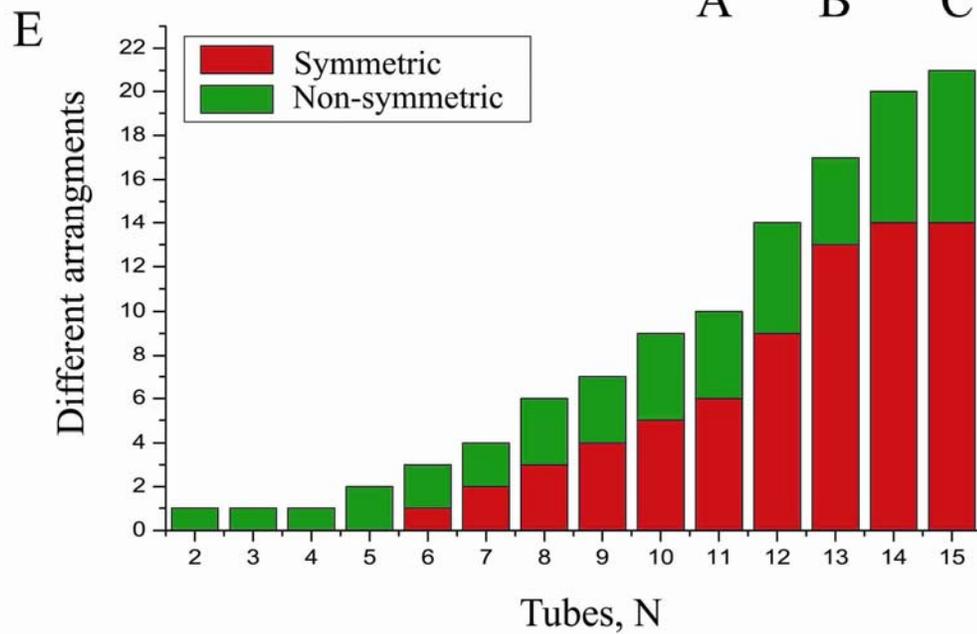